\documentstyle[12pt]{article}
\newcommand{\nsigma}{\mbox{\boldmath $\sigma$}}
\newcommand{\ntau}{\mbox{\boldmath $\tau$}}
\newcommand{\onehalf}{\frac{1}{2}}
\newcommand{\threej}[6]{ \left( \begin{array}{ccc}
                               #1 & #2 & #3 \\
                               #4 & #5 & #6 
                             \end{array}
                        \right) } 
\newcommand{\sixj}[6]{ \left\{ \begin{array}{ccc}
                                #1 & #2 & #3 \\
                                #4 & #5 & #6 
                               \end{array}
                        \right\} }

\begin{document}
\title{ {\bf Effects of state dependent correlations on nucleon
density and momentum distributions} }
\author
{F. Arias de Saavedra\\ 
{\small Departamento de Fisica
Moderna, Universidad de Granada,} \\
   {\small \sl E-18071 Granada, Spain} \\ \\
     G. Co' and M.M. Renis \\
{\small Dipartimento di Fisica, Universit\`a di Lecce and} \\
{\small INFN, Sezione di Lecce,}\\
{\small \sl I-73100 Lecce, Italy } }

\date{\mbox{ }}
\maketitle
\begin{abstract}
The proton momentum and density distributions of closed shell nuclei are
calculated with a  model treating short--range correlations up to first
order in the cluster expansion. The validity of the model is verified by
comparing the results obtained using purely scalar correlations with those
produced by finite nuclei Fermi Hypernetted Chain calculations.
State dependent correlations are used to calculate momentum and density
distributions of $^{12}$C, $^{16}$O, $^{40}$Ca, and $^{48}$Ca, and the 
effects of their tensor components are studied.
\end{abstract} 
\vskip 1.cm
PACS number(s): 21.10.Ft, 21.60.-n
\newpage
\section{Introduction}
The high precision data produced by modern electron scattering experiments
have imposed severe constraints on the validity of the models
and the theories aiming to describe the nuclear properties. Since the
beginning of the eighties, elastic scattering experiments \cite{cav82},
measuring charge distributions, and knock-out experiments \cite{qui86},
measuring spectral functions, have pointed out the difficulties of the
mean--field model in the description of the ground state of atomic nuclei.

The effects not considered in the mean--field model have been generically
named correlations. One usually distinguishes between short- and long-range
correlations.
The former ones are those acting at
short inter particle distances modifying the mean field single
particle wave functions to take care of the hard core part of
the nuclear potential. The long--range correlations are instead acting on
the full system and are produced by collective phenomena
like sound waves or surface vibrations.

The Correlated Basis Function (CBF) theory, based on Jastrow's approach
\cite{jas55}, is particularly suitable for the study of the short-range
correlations, since these correlations are explicitly included in the
definition of the many--body wave function.
Unfortunately the technical complexity of this theory has limited its
application to few--body systems and to infinite nuclear matter
\cite{cla79,pan90}. At present there are only few example of CBF
calculations in nuclei heavier than $^{4}$He. The CBF theory has been used
in Variational Monte Carlo calculations of the $^{16}$O nucleus
\cite{pie92}. Only recently the Fermi Hypernetted Chain
Theory (FHNC) has been extended to describe the ground state properties of
medium-heavy  doubly--closed shell nuclei \cite{co92,co94,ari96}.
Unfortunately these calculations are still limited to the use of central
interactions and scalar correlations. 

Because of these difficulties,
the study of short range correlations in medium--heavy nuclei has been 
done using simplified models. 
Nuclear models considering short--range correlations have been used
to analyse elastic electron scattering data  
already at the end of the sixties \cite{kha68}. 
These models are based on cluster expansion
like the CBF theory, but they retain only those terms containing a
single correlation line \cite{gau71}. The truncation of the
expansion simplifies the  calculations and it is done in a way to
conserve the normalization of the density
distribution. More recently, these simplified models have been used
to investigate also the nucleon momentum distributions
\cite{boh80,ben86,sto93}. 

The aim of the present article is twofold: first we would like to discuss
the validity of these nuclear models, and secondly we would like to
investigate the effects of the state dependent part of the correlation on
density and momentum distributions.

For these purposes we extend the model developed in Ref. \cite{co95}
to calculate both density and momentum distributions for doubly closed
shell nuclei and we compare our results with those
of finite nuclei FHNC calculations \cite{ari96}. 
After this test, using a nuclear matter correlation function \cite{wir88},
we study the effects produced by the state dependent
terms of the correlation on the proton density and the momentum
distributions of the $^{12}$C, $^{16}$O, $^{40}$Ca and $^{48}$Ca nuclei.

\section{ The model}

The basic quantity of interest for our  \ calculations  \ is the One--Body
Density Matrix (OBDM) defined as: 
\begin{equation}
\label{onebmat}
\rho({\bf r}_1, {\bf r}'_1) = \frac {A} {\cal N} \int d^3r_2 \, d^3r_3 \, ...
\, d^3r_A    \Psi^{\dagger}({\bf r}_1,{\bf r}_2,{\bf r}_3,...,{\bf r}_A)
\Psi ({\bf r}'_1,{\bf r}_2,{\bf r}_3,...,{\bf r}_A),
\end{equation}
where $\Psi$ is the nuclear ground state wave function and ${\cal N}$ 
$=<\Psi | \Psi >$. In Eq. (\ref{onebmat}) a sum on the spin and
isospin components of all the particles, particle $1$ included,
is understood.

In our model the protons and the neutrons are separately treated, therefore:
\begin{equation}
\rho({\bf r}_1, {\bf r}'_1) = \rho^p({\bf r}_1, {\bf r}'_1)
+ \rho^n({\bf r}_1, {\bf r}'_1)  
\end{equation}
where the protons and neutrons OBDMs are obtained inserting in Eq.
(\ref{onebmat}) the operators selecting the protons,
\begin{equation}
\label{q}
Q(1)= \onehalf (1+ \tau_3(1)),
\end{equation}
and the neutrons, $1-Q(1)$.

The density distribution is the diagonal part of the OBDM 
(${\bf r}_1={\bf r}'_1)$ while the momentum distribution is its Fourier
transform: 
\begin{equation}
\label{nk}
n({\bf k})=\frac {1} {A} \int d^3r_1 \int d^3r'_1 
\rho({\bf r}_1,{\bf r}'_1) e^{i {\bf k}\cdot({\bf r}_1-{\bf r}'_1)}.
\end{equation}

The OBDM is normalized as:
\begin{equation}
\label{normaa}
\int d^3 r_1 \,\, \int d^3 r'_1 \rho({\bf r}_1, {\bf r}'_1) 
\delta ({\bf r}_1 - {\bf r}'_1) = A
\end{equation}
and therefore:
\begin{equation}
\label{nknorm}
\int d^3k \, n({\bf k}) =  (2 \pi)^3 .
\end{equation}

Our model is based upon the Jastrow ansatz on the nuclear wave
function \cite{jas55}:
\begin{equation}
\Psi (1,\ldots,A)= F(1,\ldots,A) \Phi(1,\ldots,A)
\end{equation}
where $F$ is an A--body correlation operator, and $\Phi$ is a Slater
determinant built on a basis of single particle (s.p.) wave functions
generated by a one--body hamiltonian.

In the present work we shall use a spherical mean--field potential with a
spin--orbit interaction, therefore the quantum numbers characterizing the
s.p. wave functions are the principal quantum number $n$, the orbital angular
momentum $l$, the total angular momentum $j$, its third
component $m$ and  the 3rd isospin component $t$.

A usual choice of the expression of the correlation operator is based upon an
extension of the original Jastrow's ansatz to a state dependent form:
\begin{equation}
\label{jas}
F(1,\ldots,A)=S \prod^A_{j>i=1} \big[ \sum_{n=1}^M f_n (r_{ij})
O^n_{ij} \big] ,
\end{equation}
where $r_{ij}$ $=|{\bf r}_i - {\bf r}_j|$, and
we have indicated with $S$ a symmetrizer operator, with $M$ the maximum
number of the correlation channels and with $O^n_{ij}$ the operators
characterizing the various channels.  In the present work we use
correlations with 6 active channels defined as:
$O^1_{ij}=1$, $O^2_{ij}=\nsigma(i) \cdot \nsigma(j)$, 
$O^3_{ij}=\ntau(i) \cdot \ntau(j)$ 
$O^4_{ij}=\nsigma(i) \cdot \nsigma(j) \ntau(i) \cdot \ntau(j)$, 
$O^5_{ij}=S_{ij}$,
$O^6_{ij}=S_{ij} \ntau(i) \cdot \ntau(j)$, where $S_{ij}$ is the 
tensor operator:
\begin{equation}
\label{tensor}
S_{ij}= 3 \frac {(\nsigma(i) \cdot {\bf r}_{ij})(\nsigma(j) \cdot {\bf
r}_{ij}) } {(r_{ij})^2} - \nsigma(i) \cdot \nsigma(j) .
\end{equation}

The OBDM is calculated with cluster expansion techniques
applied to finite systems \cite{fan79}.
We express the correlation function as:
\begin{equation}
\label{jas1}
F(1,\ldots,A)=1+ S \prod^A_{j>i=1} \big[ \sum_{n=1}^M h_n (r_{ij})
O^n_{ij} \big] ,
\end{equation}
and perform the expansion in terms of $h_n$. 

The denominator in Eq. (\ref{onebmat}) cancels all the
unlinked diagrams of the numerator, i.e. all the diagrams which have not 
a direct link, either statistical or dynamical one, with the points
$\bf r$ and ${\bf r}'$.

The basic hypothesis of the model enters now, because of all the infinite
set of linked diagrams obtained within the cluster expansion, we retain
only those containing correlations lines up to the second order in the
correlation function $h_n$. The set of diagrams considered is shown in 
Fig.1. This is the lowest order set of correlated diagrams conserving the
normalization of the OBDM, Eq. (\ref{normaa}).

It is possible to describe this limited set of diagrams 
in terms of the correlation function $f_n$.  
In this new representation the set of diagrams to be calculated is shown in
Fig.2.

The OBDM calculated with the diagrams of Fig.2 can be expressed as:
\begin{eqnarray}
\label{rho1}
\rho^p_1({\bf r}_1, {\bf r}'_1) 
& \equiv & \rho^p_0({\bf r}_1, {\bf r}'_1) + 
A({\bf r}_1, {\bf r}'_1) - B({\bf r}_1, {\bf r}'_1) 
- C({\bf r}_1, {\bf r}'_1) + D({\bf r}_1, {\bf r}'_1)
\nonumber \\
 &=& \rho^p_0({\bf r}_1, {\bf r}'_1) 
+ \rho_0({\bf r}_1, {\bf r}'_1)
\int d^3r_2 \,\, H({\bf r}_1,{\bf r}'_1,{\bf r}_2)
\rho_0({\bf r}_2, {\bf r}_2) \nonumber \\
& & -\int d^3r_2 \rho_0({\bf r}_1, {\bf r}_2)
 \,\, H({\bf r}_1,{\bf r}'_1,{\bf r}_2)
\rho_0({\bf r}_2, {\bf r}'_1) \nonumber \\
& & -\int d^3r_2 \int d^3r_3  \rho_0({\bf r}_1, {\bf r}_2)
\rho_0({\bf r}_2, {\bf r}'_1) \rho_0({\bf r}_3, {\bf r}_3)
 \,\, H({\bf r}_2,{\bf r}_2,{\bf r}_3)
\nonumber \\
& & +\int d^3r_2 \int d^3r_3  \rho_0({\bf r}_1, {\bf r}_2)
\rho_0({\bf r}_2, {\bf r}_3) \rho_0({\bf r}_3, {\bf r}'_1)
 \,\, H({\bf r}_2,{\bf r}_2,{\bf r}_3) ,
\end{eqnarray}
where,
like in Eq. (\ref{onebmat}), the sum on spin and isospin components
is understood.

In the above expression we have used the uncorrelated OBDM 
defined in terms of the s.p. wave functions as: 
\begin{equation}
\label{rhou}
\rho_0({\bf r}_1, {\bf r}_2) = \sum_{nljmt} \phi^{*t}_{nljm}({\bf r}_1)
\phi^t_{nljm}({\bf r}_2). 
\end{equation}
In the diagrams of Fig.2 the $\rho_0({\bf r}_1, {\bf r}_2)$ is
represented by an oriented line. The dashed lines represent the
dynamical correlations  expressed in Eq. (\ref{rho1}) by the coefficients 
$H({\bf r}_1,{\bf r}_2,{\bf r}_3)$ defined as: 
\begin{equation}
\label{correlation}
H({\bf r}_1,{\bf r}_2,{\bf r}_3)
= \Big( \sum_{p=1}^6 f_p (r_{13}) O^p_{13} \Big)
Q(1)
\Big( \sum_{q=1}^6 f_q (r_{23}) O^q_{13} \Big),
\end{equation}
with $Q(1)$  defined by Eq. (\ref{q}).

It is easy to see that
the set of considered diagrams conserves the density normalization. Because
of the property of the uncorrelated density
\begin{equation}
\int d^3r_j \,\, \rho_0({\bf r}_i,{\bf r}_j) 
\rho_0 ({\bf r}_j,{\bf r}_k) = \rho_0 ({\bf r}_i,{\bf r}_k),
\end{equation}
where we have also summed on the spin and isospin coordinates of
the particle $j$, we obtain the following relations among the diagrams 
of Fig.2:
\begin{equation}
\int d^3r_1 \int d^3r'_1 \,\, A({\bf r}_1,{\bf r}'_1) \,\,
\delta ({\bf r}_1-{\bf r}'_1) =
\int d^3r_1 \int d^3r'_1 \,\, C({\bf r}_1,{\bf r}'_1) \,\,
\delta ({\bf r}_1-{\bf r}'_1) 
\end{equation}
\begin{equation}
\int d^3r_1 \int d^3r'_1 \,\, B({\bf r}_1,{\bf r}'_1) \,\,
\delta ({\bf r}_1-{\bf r}'_1) =
\int d^3r_1 \int d^3r'_1 \,\, D({\bf r}_1,{\bf r}'_1) \,\,
\delta ({\bf r}_1-{\bf r}'_1) 
\end{equation}
and therefore:
\begin{equation}
\int d^3r_1 \int d^3r'_1 \,\, \rho^p_1 ({\bf r}_1,{\bf r}'_1) \,\,
\delta ({\bf r}_1-{\bf r}'_1) =
\int d^3r_1 \int d^3r'_1 \,\, \rho^p_0 ({\bf r}_1,{\bf r}'_1) \,\,
\delta ({\bf r}_1-{\bf r}'_1) = Z
\end{equation}

In the evaluation of the spin and isospin traces of the correlation kernels
of Eq. (\ref{correlation}),
we found convenient to separate the various terms with respect to
their isospin dependence and we obtain four terms:
\begin{eqnarray}
& & H({\bf r}_1,{\bf r}'_1,{\bf r}_2) = \nonumber \\
& & \Big[ f_1g_1 + 3f_2g_2 + 6f_5g_5 +  
\Bigl( f_1g_2 + f_2g_1 - 2f_2g_2 + 2f_5g_5 \Bigr)\nsigma(1) \cdot
\nsigma(2) + \nonumber \\
& & 
\Bigl(f_1g_5 + f_5g_1 + f_2g_5 + f_5g_2 - 2 f_5g_5 \Bigr) S_{12} 
\Big] Q(1)
+ \nonumber \\
& & \Big[ f_3g_1 + 3f_4g_2 + 6f_6g_5 +
\Bigl(f_3g_2 +f_4g_1 - 2f_4g_2 + 2f_6g_5 \Bigr)
\nsigma(1) \cdot \nsigma(2)
+ \nonumber \\ 
& & 
\Bigl( f_3g_5 + f_6g_1 + f_4g_5 + f_6g_2 - 2f_6g_5 \Bigr) S_{12} 
\Big]\ntau(1) \cdot \ntau(2) Q(1) +
\nonumber \\ 
& & \Big[ f_1g_3 + 3f_2g_4 + 6f_5g_6 +
\Bigl( f_1g_4 + f_2g_3 - 2f_2g_4+ 2f_5g_6 \Bigr)
\nsigma(1) \cdot
\nsigma(2) + \nonumber \\ 
& & \Bigl(f_1g_6 + f_5g_3 + f_2g_6 + f_5g_4 - 2f_5g_6 \Bigr)
S_{12}  \Big] Q(1) \ntau(1) \cdot \ntau(2)  +
\nonumber \\
& &\Big[ f_3g_3 + 3f_4g_4 + 6f_6g_6 + 
\Bigl( f_3g_4 + f_4g_3 - 2f_4g_4 +
2f_6g_6 \Bigr) \nsigma(1) \cdot \nsigma(2) +  \nonumber \\
& & 
\label{acca}
\Bigl(f_3g_6 + f_6g_3 + f_4g_6 + f_6g_4 - 2f_6g_6
\Bigr) S_{12} \Big]
\ntau(1) \cdot \ntau(2) Q(1) \ntau(1) \cdot \ntau(2)
\end{eqnarray}
In the above expression to simplify, the writing, we used the symbol $f_n$
for the correlation functions depending from ${\bf r}_1$ and ${\bf r}_2$,
and $g_n$ for those depending from ${\bf r}'_1$ and ${\bf r}_2$. Clearly,
in the case ${\bf r}_1={\bf r}'_1$  we have $f_n=g_n$.

The calculation continues inserting the above expression in Eq.
(\ref{rho1}) and evaluating the spin and isospin traces.
The traces of the isospin dependent part of each term are given in Tab.1.
We have explicitly verified that the spin traces of all the terms
depending from the tensor operator are zero. This is due to the definition
of the tensor operator, Eq. (\ref{tensor}), to the spherical symmetry
of the problem and to the saturation of the spin of all the single particle
wave functions.

In order to express the final result we used the quantity  
$\rho_0^{s1,s2,t}({\bf r}_1,{\bf r}_2)$ 
defined by the relation:
\begin{equation}
\rho_0({\bf r}_1,{\bf r}_2)=
\sum_{t=p,n} \sum_{s1s2} \rho_0^{s1,s2,t}({\bf r}_1,{\bf r}_2)
\chi_{s1}^\dagger(1) \chi_{s2}(2)
\chi_{t}^\dagger(1) \chi_{t}(2),
\end{equation}
where we have indicated with $\chi$ the spin and isospin wave functions.
Together with the above definition we define as well:
\begin{eqnarray}
\rho_0^t({\bf r}_1,{\bf r}_2) &=& 
\sum_s \rho_0^{s,s,t}({\bf r}_1,{\bf r}_2) \\
\rho_0^t({\bf r}_1) &=& \rho_0^t({\bf r}_1,{\bf r}_1).
\end{eqnarray}

Using the quantities defined above we
express the four terms composing $\rho^p_1$, Eq.
(\ref{rho1}), as:
\begin{eqnarray}
\label{primo}
A({\bf r}_1,{\bf r}'_1) &=&\rho_0^p({\bf r}_1,{\bf r}'_1)
\int d^3r_2 \big[ \rho_0^p({\bf r}_2) G1({\bf r}_1,{\bf r}'_1,{\bf r}_2)
           +\rho_0^n({\bf r}_2) G2({\bf r}_1,{\bf r}'_1,{\bf r}_2) \big] 
\nonumber \\
&+&4 \rho_0^n({\bf r}_1,{\bf r}'_1)
\int d^3r_2 \rho_0^p({\bf r}_2) G3({\bf r}_1,{\bf r}'_1,{\bf r}_2)
\\
%
& & \nonumber \\
%
\label{secondo}
B({\bf r}_1,{\bf r}'_1) &=& \int d^3r_2   \Bigl\{  \sum_{s1\,s2} 
\rho_0^{s1,s2,p} ({\bf r}_1,{\bf r}_2)
\rho_0^{s2,s1,p} ({\bf r}_2,{\bf r}'_1) \,\,
G4({\bf r}_1,{\bf r}'_1,{\bf r}_2)  
\nonumber \\ 
&+& 
2 \rho_0^p ({\bf r}_1,{\bf r}_2) \rho_0^p ({\bf r}_2,{\bf r'_1})
\,\, G5({\bf r}_1,{\bf r}'_1,{\bf r}_2) 
\nonumber \\
&+& \sum_{s1\,s2} 
2\rho_0^{s1,s2,p} ({\bf r}_1,{\bf r}_2)
\rho_0^{s2,s1,n} ({\bf r}_2,{\bf r}'_1) \,\,
G6({\bf r}_1,{\bf r}'_1,{\bf r}_2)  
\nonumber \\
&+& 
4 \rho_0^p ({\bf r}_1,{\bf r}_2) \rho_0^n ({\bf r}_2,{\bf r'_1})
\,\, G7({\bf r}_1,{\bf r}'_1,{\bf r}_2) 
\nonumber \\
&+& \sum_{s1\,s2} 
2\rho_0^{s1,s2,n} ({\bf r}_1,{\bf r}_2)
\rho_0^{s2,s1,p} ({\bf r}_2,{\bf r}'_1) \,\,
G8({\bf r}_1,{\bf r}'_1,{\bf r}_2)  
\nonumber \\
&+& 
4 \rho_0^n ({\bf r}_1,{\bf r}_2) \rho_0^p ({\bf r}_2,{\bf r'_1})
\,\, G9({\bf r}_1,{\bf r}'_1,{\bf r}_2) 
\Bigr\}
\\
%
& & \nonumber \\
%
\label{terzo}
C({\bf r}_1,{\bf r}'_1) &=& \int d^3r_2 \int d^3r_3 \sum_{s1\,s2}
\rho_0^{s1,s2,p} ({\bf r}_1,{\bf r}_2)
\rho_0^{s2,s1,p} ({\bf r}_2,{\bf r}'_1) \,\,
\nonumber \\
& &
\Bigl[ \rho_0^p ({\bf r}_3) G1({\bf r}_2,{\bf r}_2,{\bf r}_3)  
+\rho_0^n ({\bf r}_3) \left( G2({\bf r}_2,{\bf r}_2,{\bf r}_3) +
4 G3({\bf r}_2,{\bf r}_2,{\bf r}_3) \right) \Bigr]
\\
& & \nonumber \\
%
\label{quarto}
D({\bf r}_1,{\bf r}'_1) &=& \int d^3r_2 \int d^3r_3 \Bigl\{
\sum_{s1\,s2\,s3}
\rho_0^{s1,s2,p} ({\bf r}_1,{\bf r}_2)
\rho_0^{s3,s1,p} ({\bf r}_3,{\bf r}'_1) \,\,
\nonumber \\
& &
\hspace*{-0.3cm} \Bigl[  \rho_0^{s2,s3,p} ({\bf r}_2,{\bf r}_3)
G4({\bf r}_2,{\bf r}_2,{\bf r}_3)  +
2 \rho_0^{s2,s3,n} ({\bf r}_2,{\bf r}_3)
\left( G6({\bf r}_2,{\bf r}_2,{\bf r}_3) +
 G8({\bf r}_2,{\bf r}_2,{\bf r}_3) \right) \Bigr]
\nonumber \\
&+& \sum_{s1\,s2}
\rho_0^{s1,s2,p} ({\bf r}_1,{\bf r}_2)
\rho_0^{s2,s1,p} ({\bf r}_3,{\bf r}'_1) \,\,
\nonumber \\
& &
2 \Bigl[  \rho_0^p ({\bf r}_2,{\bf r}_3)
G5({\bf r}_2,{\bf r}_2,{\bf r}_3)  +
2 \rho_0^n ({\bf r}_2,{\bf r}_3)
 \left( G7({\bf r}_2,{\bf r}_2,{\bf r}_3) +
 G9({\bf r}_2,{\bf r}_2,{\bf r}_3) \right) \Bigr]
\Bigr\}
\end{eqnarray}
The explicit expression of the terms $G$ containing the correlation
functions are given in the Appendix.

The calculation of the momentum and density
distributions of doubly--magic nuclei has been done by exploiting the
spherical symmetry of the problem to perform an analytical integration on
the angular variables.
For this purpose we have expressed the s.p. wave functions in spherical
coordinates and we have performed a multipole expansion of the
functions $G$ containing the correlations. In the appendix we present the
basic points of the calculation and we give
the final expressions used to evaluate the diagrams $A$, $B$, $C$
and $D$. The details of the calculation are presented in Ref. \cite{ren95}.

\section{Results}

In our calculations the s.p. wave functions have been generated by a spherical
Woods--Saxon  well of the form:
\begin{equation}
\label{wood}
V({\bf r})={-V_0 \over 1+e^{(r-R)/a} }+
\left( { \hbar \over m_\pi c} \right)^2 {1 \over r} 
{d \over dr} \left( {-V_{ls} \over 1+e^{(r-R)/a} } \right) \;
{\bf l} \cdot  {\mbox{\boldmath $\sigma$}} + V_{Coul},
\end{equation}
where $m_\pi$ is the pion mass.

For each nucleus considered, we have taken from the literature \cite{co84}
the parameters of the potential (see Tab.2). These parameters
have been fixed to reproduce the s.p.
energies around the Fermi surface and the root mean squared charge radii.

We tested the validity of our model by comparing our results with 
those of FHNC calculations performed with the
same inputs. We used two kinds of correlation functions. A first
one has a gaussian functional dependence from the inter particle
distance. The second one has been obtained in Ref. \cite{ari96} by
solving the variational equations for the mean value of the 
hamiltonian up to the second order (Euler correlations). 

The correlations used for the $^{12}$C and $^{48}$Ca nuclei are
shown in Fig.3. The dashed lines represent the gaussian
correlations, while the full lines show the
Euler correlations. These correlations have been fixed
in Ref. \cite{ari96} by minimizing the nuclear binding energies for the
Afnan and Tang S3 interaction \cite{afn68}. The Euler correlations
show an overshooting of the asymptotic value in the region between
1 and 2 fm.

In Fig.4 we compare the momentum and density
distributions calculated within the FHNC theory \cite{ari96} (dashed
lines) and those obtained with the present model (full lines).
The agreement between the two calculations is quite good, 
independently from the correlation function used.

Confident of the validity of our model we use the nuclear matter
state dependent correlation function of Ref. \cite{wir88} to
calculate momentum and density distributions in $^{12}$C, $^{16}$O,
$^{40}$Ca and $^{48}$Ca.

We show in Fig.5 the various terms of the correlation as a function of the
inter particle distance. The dominant term of the correlation is the scalar
channel.  The tensor term ($f_5$) is extremely small and the tensor--isospin
term ($f_6$) is peaked at an inter particle distance of about 1 fm.

In Figs.6 and 7 we show the proton density and momentum
distributions for the four nuclei considered calculated by switching on
and off the various channels of the correlation function. The dotted
lines represent the uncorrelated results. The results of the
calculations performed with the correlation containing all the six
channels are shown by the full lines. The dashed lines have been obtained
using correlations with only the scalar channel active, while the
dashed--dotted lines  with correlations without the
two tensor channels.

The results of Fig.6
show the same behaviour for each nucleus considered. In
the nuclear interior, the correlated density distributions are
smaller than the uncorrelated ones. We notice that the
curves obtained with purely scalar correlations are the more
distant ones from the uncorrelated results. The inclusion of the
other correlation channels reduces these differences.

The momentum distributions (Fig.7) show a well known
behaviour. Correlated and uncorrelated results are practically the same up
to momentum values of about 1.8--2.0 fm$^{-1}$, but they clearly 
separate at higher values. The uncorrelated distributions descend very
rapidly while the correlated distributions are order of magnitude larger.

In Fig.7 we observe that the full lines and dashed lines, 
obtained with correlations containing only the scalar term,
are well separated in the high momentum region.
On the contrary, the full and dashed--dotted lines, the last ones obtained
using correlations without the tensor channels,
differ only in a small momentum region around 2.0 fm$^{-1}$. 
This fact leads us to think that the effects of the tensor correlations are
well localized in momentum space.

To investigate better this issue we have performed calculations of the
momentum distributions using correlations with only the 
tensor channels active. The result for the $^{12}$C nucleus are
presented in Fig.8 (dotted line) and is compared with the momentum
distribution calculated with the full correlation (full line) and with that
obtained with central and tensor terms only (dashed line). 

We observe that the filling of the dip around 2 $fm^{-1}$ is
produced by the sum of the effects of the scalar and tensor
components. The other central channels are responsible of the
increase of the tail of the distribution at higher momentum values.

\section {Conclusions}
We have developed a nuclear model to describe density and momentum
distributions of doubly--closed shell nuclei explicitly considering the
short-range correlations.

This model is based upon the cluster expansion of the CBF theory,
but it retains only the set of lowest order diagrams which allows
for the conservation of the number of particles. 

The model has been tested against the results obtained by finite
nuclei FHNC calculations performed with the same inputs. The
agreement between our results and those of the more elaborated
theory is very good.

We have calculated momentum and density distributions of four
doubly closed shell nuclei with state dependent correlations taken
from nuclear matter FHNC calculations.

We found that the major effects of the correlations in both density
and momentum distributions are produced by the scalar part of the
correlations. The effect of the correlation functions used in these
calculations is a general lowering of the 
density distributions in the interior region and an increasing of the
momentum distribution at high momentum values. 

The effect of the state dependent terms of the correlation on the
density distribution is of opposite sign with respect to the effect
of the scalar term only. The results obtained with the complete
correlation are closer to the uncorrelated results than those
obtained with the scalar term only.
The situation is reversed in the momentum distribution case. 

The tensor correlation term produce effects which seem to be rather
localized in momentum space. In the momentum distributions, the
presence of the tensor correlations is noticeable in a small momentum
region around the point where the uncorrelated distributions
separate from the correlated ones.

Before concluding we would like to make two comments about the
limitations of the model with respect to FHNC calculations.

A first limitation is of practical type. Contrary to what happens
in the FHNC case, the numerical effort to perform our calculations
grows rapidly with the number of single particle states. This makes
the calculation of the $^{208}$Pb momentum distribution extremely
heavy from the numerical point of view.

The second limitation has a theoretical aspect common to all the
models of the same kind. In the FHNC theory the
correlation function and the s.p. basis are related through the nuclear
hamiltonian by the variational principle. In our model they are two
different input parameters in principle arbitrary. 

We have performed our calculations using reasonable s.p. bases and
reasonable correlations. The s.p. bases have been taken from
literature where their parameters have been fixed in order to
reproduce some nuclear properties within a mean--field model. This
could mean that some effect of the short range correlations we
would like to describe have already been averaged out by this
procedure. The correlation function used has been obtained by a
minimization procedure done in nuclear matter. We show
in Fig.3 that finite nuclei FHNC minimizations seems to prefer
correlation functions with an overshooting in the region between 1
and 2 fm, but this is not the characteristic shown by the scalar
term of the nuclear matter correlation function.

Because of these theoretical limitations, we believe that
at the present stage the comparison of the results of
our model with experimental data is not very meaningful. The
validity of our work lies in the evaluation of the relative effects
produced by the various correlation channels.

\vskip 1.cm
\noindent
{\bf Acknowledgments}
This work has been partially supported by the agreement CYCIT--INFN and the
Junta de Andalucia.

\section* {Appendix}

In this appendix we present the expressions of the equations used to
calculate the  diagrams of Fig.1.

The functions $G$ of Eqs. (\ref{primo}) are defined as:
\begin{eqnarray}
G1 &=& (f_1+f_3)(g_1+g_3)+3(f_2+f_4)(g_2+g_4)
         +6(f_5+f_6)(g_5+g_6) \\
\nonumber  \\
G2 &=& (f_1-f_3)(g_1-g_3)+3(f_2-f_4)(g_2-g_4)
         +6(f_5-f_6)(g_5-g_6) \\
\nonumber  \\
G3 &=& f_3 g_3+3f_4 g_4 + 6 f_6 g_6 \\
\nonumber  \\
G4 
   &=& (f_1+f_3)(g_1+g_3)+[\onehalf(f_2+f_4)-f_1-f_3](g_2+g_4)
\nonumber \\
   &+& (f_2+f_4) [\onehalf(g_2+g_4)-g_1-g_3]
         +4(f_5+f_6)(g_5+g_6) \\
\nonumber \\
G5 
   &=&
(f_1-f_2+f_3-f_4)(g_2+g_4)+(f_2+f_4)(g_1-g_2+g_3-g_4) 
\nonumber \\
&+& 2(f_5+f_6)(g_5+g_6)\\
\nonumber \\
G6 
   &=&
(f_1-f_3)(g_3-g_4)+(f_2-f_4)(5g_4-g_3) + 4(f_5-f_6)g_6 \\
\nonumber \\
G7 
   &=&
(f_2-f_4)g_3+(f_1-2f_2-f_3+2f_4)g_4 + 2(f_5-f_6)g_6 \\
\nonumber \\
G8 
   &=&
(f_3-f_4)(g_1-g_3)+(5f_4-f_3)(g_2-g_4) + 4f_6(g_5-g_6) \\
\nonumber \\
G9 
   &=&
f_3(g_2-g_4)+f_4(g_1-2g_2-g_3+2g_4) + 2f_6(g_5-g_6) \\
\nonumber \\
\end{eqnarray}

To describe the closed shell nuclei we have used a set
of s.p. wave functions of the form:
\begin{equation}
\label{sp}
\phi^t_{nljm}({\bf r}_i)=
R^t_{nlj}(r_i) \sum_{\mu,s} <l \mu 1/2 s | j m > Y_{l\mu}(\Omega_i)
\chi_s(i)\chi_t(i).
\end{equation}
In the above equation $\Omega_i$ indicates the angular coordinates, $l$ and
$j$ the orbital and total angular momentum respectively, $R_{nlj}(r)$ the
radial part of the wavefunction, $Y_{l\mu}$ the spherical harmonic and 
$<l \mu 1/2 s | j m >$ the Clebsch--Gordan coefficient.

In this basis the uncorrelated proton OBDM, the first term of Eq.
(\ref{rho1}) is : 
\begin{equation}
\rho^p_0({\bf r}_1,{\bf r}_2)=
\frac{1}{4\pi} \sum_{nlj} R^p_{nlj}(r_1)R^p_{nlj}(r_2)
(2j+1) P_l(cos \theta_{12})
\end{equation}
where $P_l$ is a the Legendre polynomial and $\theta_{12}$ is the angle
between ${\bf r}_1$ and ${\bf r}_2$.

The contribution of the other terms has been calculated using the
expression (\ref{sp}) of the s.p. wave functions and performing a multipole
expansion of each correlation function $f_n$ and $g_n$ of the $G$ functions:
\begin{equation}
f_\alpha ({\bf r}_1,{\bf r}_2) = \sum_L f^\alpha_L(r_1,r_2) 
P_L(\cos \theta_{12}) 
\end{equation}
where $\alpha=1,...,6$.

The contribution of the diagrams $A,B,C$ and $D$ has beeen calculated
expanding in multipole each term composing the expressions
from (\ref{primo}) to (\ref{quarto}). Each of
these terms contains the product between the correlation functions $f_n$ and
$g_n$. In the diagram $A$ we found terms of the form:
\begin{eqnarray}
& &\rho_0^{t_1} ({\bf r}_1,{\bf r}_1') 
 \int d^3r_2 \rho_0^{t_2} (r_2) f_\alpha(r_{12}) g_\beta(r_{1'2}) 
 =  \sum_{n_1,l_1,j_1} (2 j_1+1) R_{n_1,l_1,j_1}^{t_1} (r_1)
R_{n_1,l_1,j_1}^{t_1} (r_1') \nonumber \\
& &  \sum_{l_2,l_3}  \frac {2l_3+1}{2l_2+1} 
\threej{l_1}{l_3}{l_2}{0}{0}{0}^2
P_{l_3} (\cos \theta_{11'}) \int_0^\infty dr_2 r_2^2
\rho_0^{t_2} (r_2) f^\alpha_{l_2}(r_1,r_2) g^\beta_{l_2}(r_1',r_2)
\end{eqnarray}
In the diagram $B$ we calculate terms of the form:
\begin{eqnarray}
& &\int d^3r_2 \rho_0^{t_1} ({\bf r}_1,{\bf r}_2)
\rho_0^{t_2} ({\bf r}_2,{\bf r}_1')  f_\alpha(r_{12})
g_\beta (r_{1'2}) =   \frac 1 {4 \pi} \sum_{n_1,l_1,j_1 \atop
n_2,l_2,j_2}  (2 j_1+1)(2 j_2+1)
\nonumber \\
& & R_{n_1,l_1,j_1}^{t_1} (r_1)
R_{n_2,l_2,j_2}^{t_2} (r_1')  \sum_{l_3,l_4,l_5}  (2l_5+1)
\threej{l_1}{l_5}{l_3}{0}{0}{0}^2
\threej{l_2}{l_5}{l_4}{0}{0}{0}^2 
\nonumber \\
& &\int_0^\infty dr_2 r_2^2
R_{n_1,l_1,j_1}^{t_1} (r_2) R_{n_2,l_2,j_2}^{t_2} (r_2)
f^\alpha_{l_3}(r_1,r_2) g^\beta_{l_4}(r_1',r_2)
\,\,P_{l_5} (\cos \theta_{11'}) 
\end{eqnarray}
and
\begin{eqnarray}
& &\sum_{s_1,s_2}\int d^3r_2 \rho_0^{s_1 s_2 t_1} ({\bf
r}_1,{\bf r}_2) \rho_0^{s_2 s_1 t_2} ({\bf r}_2,{\bf r}_1')  
f_\alpha(r_{12}) g_\beta(r_{1'2})  =  
\nonumber \\
& &- \frac 1 {4 \pi} \sum_{n_1,l_1,j_1 \atop n_2,l_2,j_2} 
\sqrt{(2l_1+1)(2l_2+1)} (2 j_1+1)(2 j_2+1)  R_{n_1,l_1,j_1}^{t_1}
(r_1) R_{n_2,l_2,j_2}^{t_2} (r_1')  
\nonumber \\
& &\sum_{l_3,l_4,l_5,l_6}    (-1)^{l_5-l_6} 
\frac {1+(-1)^{l_1+l_2+l_5}} 2 (2l_5+1)(2l_6+1)
 \nonumber \\
& &  \int_0^\infty dr_2 r_2^2
R_{n_1,l_1,j_1}^{t_1} (r_2) R_{n_2,l_2,j_2}^{t_2} (r_2)
f_{l_3}^\alpha (r_1,r_2) g_{l_4}^\beta(r_1',r_2)
\nonumber \\
& &
\threej{j_2}{j_1}{l_5}{1/2}{-1/2}{0} \threej{l_3}{l_5}{l_4}{0}{0}{0}
\threej{l_2}{l_6}{l_4}{0}{0}{0} \threej{l_1}{l_6}{l_3}{0}{0}{0}
\nonumber \\ 
& & 
\sixj{l_1}{l_2}{l_5}{j_2}{j_1}{1/2}
\sixj{l_2}{l_4}{l_6}{l_3}{l_1}{l_5} 
\,\, P_{l_6} (\cos \theta_{11'}) \nonumber
\end{eqnarray}
In the diagrams $C$ and $D$ the functions describing the
correlations depend only on two radial coordinates, therefore
we can perform the multipole expansion of their product. So we shall
define
\begin{equation}
h_{\alpha \beta} (r_{23}) =  f_\alpha(r_{23})  f_\beta(r_{23})
\end{equation}

In the diagram $C$ we obtain terms of the form:
\begin{eqnarray}
& &\sum_{s_1,s_2}\int d^3r_2 d^3r_2 \rho_0^{s_1 s_2 p} 
({\bf r}_1,{\bf r}_2) \rho_0^{s_2 s_1 p} ({\bf r}_2,{\bf r}_1')  
\rho_0^{t} (r_3)h_{\alpha \beta}(r_{23}) =  
\nonumber \\
& &
\sum_{n_1,l_1,j_1 \atop n_2} (2 j_1+1) 
R_{n_1,l_1,j_1}^{p} (r_1)
R_{n_2,l_1,j_1}^{p} (r_1') 
\int_0^\infty dr_2 r_2^2
R_{n_1,l_1,j_1}^{p} (r_2) R_{n_2,l_1,j_1}^{p} (r_2)
\nonumber \\
& & 
\int_0^\infty dr_3 r_3^2 \rho_0^{t} (r_3) 
h^{\alpha \beta}_{0}(r_2,r_3) P_{l_1}
(\cos\theta_{11'})   ,
\end{eqnarray}
while the diagram $D$ we calculate terms of the form:
\begin{eqnarray}
& &\sum_{s_1,s_2}\int d^3r_2 d^3r_3 \rho_0^{s_1 s_2 p} 
({\bf r}_1,{\bf r}_2) \rho_0^{t} ({\bf r}_2,{\bf r}_3)
\rho_0^{s_2 s_1 p} ({\bf r}_3,{\bf r}_1')  
h_{\alpha \beta} (r_{23}) =  
\nonumber \\
& & \frac 1 {4 \pi} \sum_{n_1,l_1,j_1 \atop n_2,l_2,j_2 \  n_3} 
(2 j_1+1)(2 j_2+1) 
R_{n_1,l_1,j_1}^{p} (r_1)
R_{n_3,l_1,j_1}^{p} (r_1') P_{l_1} (\cos \theta_{11'}) 
\nonumber \\
& &  \sum_{l_4} \threej{l_1}{l_2}{l_4}{0}{0}{0}^2 
B(r_1,r_1') 
\end{eqnarray}
and
\begin{eqnarray}
& &\sum_{s_1,s_2,s_3}\int d^3r_2 d^3r_3 \rho_0^{s_1 s_2 p} 
({\bf r}_1,{\bf r}_2) \rho_0^{s_2 s_3 t} ({\bf r}_2,{\bf r}_3)
\rho_0^{s_3 s_1 p} ({\bf r}_3,{\bf r}_1')  
h_{\alpha \beta} (r_{23})  =  
\nonumber \\
& &  \frac 1 {4 \pi} \sum_{n_1,l_1,j_1 \atop n_2,l_2,j_2 \  n_3} 
(2 j_1+1)(2 j_2+1) 
R_{n_1,l_1,j_1}^{p} (r_1)
R_{n_3,l_1,j_1}^{p} (r_1') P_{l_1} (\cos \theta_{11'}) 
\nonumber \\
& &  
\sum_{l_4} \frac {1+(-1)^{l_1+l_2+l_4}} 2
\threej{j_2}{j_1}{l_4}{1/2}{-1/2}{0}^2  B(r_1,r_1')
\end{eqnarray}

where we have defined:

\begin{eqnarray}
B(r_1,r_1')  &=&  \int_0^\infty dr_2 r_2^2
R_{n_1,l_1,j_1}^{p} (r_2) R_{n_2,l_2,j_2}^{t} (r_2)
\nonumber \\
& &\int_0^\infty dr_3 r_3^2 R_{n_2,l_2,j_2}^{t} (r_3) 
R_{n_3,l_1,j_1}^{p} (r_3) h^{\alpha \beta}_{l_4}(r_2,r_3).
\end{eqnarray}

\newpage
\section*{Tables}

\begin{center}
\begin{tabular}{|c|cc|}
\hline
     &  $A$  & $C$ \\
\hline
& &  \\
$<Q(1)>$   & $\delta_{t1,p}$ & $\delta_{t1,p}\delta_{t2,p}$ \\
$<\ntau(1) \cdot \ntau(2) Q(1) >$   
         & $\delta_{t1,p}(2\delta_{t2,p}-1)$ 
         & $\delta_{t1,p}\delta_{t2,p}(2\delta_{t3,p}-1)$ \\
$<Q(1) \ntau(1) \cdot \ntau(2)>$   & $\delta_{t1,p}(2\delta_{t2,p}-1)$ 
         & $\delta_{t1,p}\delta_{t2,p}(2\delta_{t3,p}-1)$ \\
$<\ntau(1) \cdot \ntau(2) Q(1) \ntau(1) \cdot \ntau(2)>$   
         & $(2\delta_{t2,p}-\delta_{t1,p})^2$ 
         & $\delta_{t1,p}\delta_{t2,p}(5-4\delta_{t3,p})$ \\
& & \\
\hline
   &  $B$  & $D$ \\
\hline
& & \\
$<Q(1)>$    & $\delta_{t1,p}\delta_{t2,p}$
            & $\delta_{t1,p}\delta_{t2,p}\delta_{t3,p}$  \\
$<\ntau(1) \cdot \ntau(2) Q(1) >$ 
       & $\delta_{t1,p}(2-\delta_{t2,p})$
       & $\delta_{t1,p}\delta_{t3,p}(2-\delta_{t2,p})$ \\
$< Q(1) \ntau(1) \cdot \ntau(2)>$ 
       & $\delta_{t2,p}(2-\delta_{t1,p})$
       & $\delta_{t1,p}\delta_{t3,p}(2-\delta_{t2,p})$ \\
$<\ntau(1) \cdot \ntau(2) Q(1) \ntau(1) \cdot \ntau(2)>$ 
       &$5\delta_{t1,p}\delta_{t2,p}-2\delta_{t1,p}-2\delta_{t2,p}$
       & $\delta_{t1,p}\delta_{t3,p}(5\delta_{t2,p}-4)$ \\
& &  \\
\hline
\end{tabular}
\end{center}
{\bf Table 1.} Isospin traces of eq. (\ref{acca}) for each diagram considered.

\vskip 2.cm
\begin{center}
\begin{tabular}{|c|c|cccc|}
\hline
         &  &  $V_0$ & $V_{LS}$  & $R$ & $a$ \\
\hline
$^{12}$C   & p  & 62.00 & 3.20 & 2.86 & 0.57   \\
           & n  & 60.00 & 3.15 & 2.86 & 0.57   \\
\hline
$^{16}$O   & p  & 52.50 & 7.00 & 3.20 & 0.53   \\
           & n  & 52.50 & 6.54 & 3.20 & 0.53   \\
\hline
$^{40}$Ca  & p  & 57.50 & 11.11 & 4.10 & 0.53   \\
           & n  & 55.00 & 8.50 & 4.10 & 0.53   \\
\hline
$^{48}$Ca  & p  & 59.50 & 8.55 & 4.36 & 0.53   \\
           & n  & 50.00 & 7.74 & 4.36 & 0.53   \\
\hline
\end{tabular}
\end{center}
{\bf Table 2.} Coefficients of the Woods--Saxon potential eq. (\ref{wood}). 

\newpage
{\bf Figure captions}

\vskip 1. cm

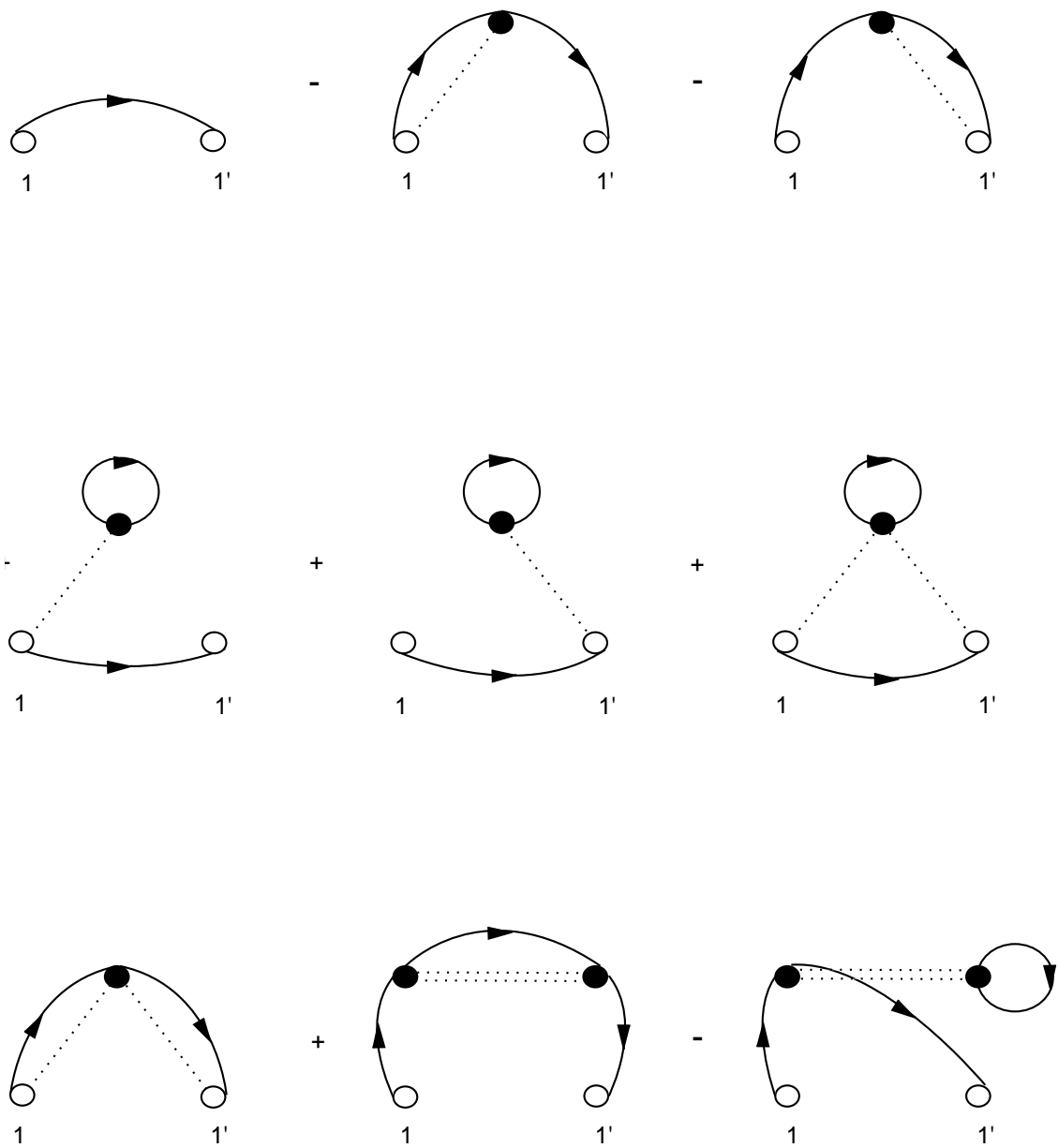
{\bf Fig.1.} Set of diagrams considered in our model. The oriented lines
represent the uncorrelated OBDM (Eq. \ref{jas1}) and the dotted lines
represent the correlation functions $h_n$ (Eq. \ref{rhou}).

\vskip 1. cm
{\bf Fig.2.} Set of diagrams considered in our model. This set of diagrams
corresponds to that of Fig.1. In this figure the dashed lines indicate
the dynamical correlations $f_n$ (Eq. \ref{jas}).

\vskip 1. cm
{\bf Fig.3.} Correlation functions used in 
the $^{12}$C and $^{48}$Ca calculation of Fig.4. The dashed
lines show the gaussian correlations and the full lines the correlations
obtained with a Euler minimization procedure.

\vskip 1.cm
{\bf Fig.4.} Comparison between density (upper panels) and momentum
distributions (lower panels) calculated with our model (dashed lines) and
with a full FHNC calculation \cite{ari96}. 
We have added a factor 1.2 to the densities calculated with the
Euler correlations and we have been multiplied the Euler
momentum distributions by a factor 10.

\vskip 1.cm
{\bf Fig.5.} State dependent correlation terms as a function of the
interparticle distance \cite{wir88}. See Eq. (\ref{jas}) for the meaning 
of the various lines.

\vskip 1.cm
{\bf Fig.6.} Proton density distributions calculated with the correlation
function of Fig.5. The dotted lines show the uncorrelated results. 
The full lines have been obtained with the full correlations. 
The results of the calculations performed without the two tensor components of
the correlations are shown by the dashed dotted lines, while the dashed lines
show the results obtained with only the scalar term of the correlation
($f_1$).

\vskip 1.cm
{\bf Fig.7.} Proton momentum distributions. The meaning of the various lines
is the same as in Fig.5.

\vskip 1.cm
{\bf Fig.8.} Proton momentum distribution of $^{12}$C. The full line
correspond to the full calculation (the same as in Fig.7). The dotted line
has been obtained using only the two tensor components of the correlation
and the dashed line with the scalar plus tensor components.

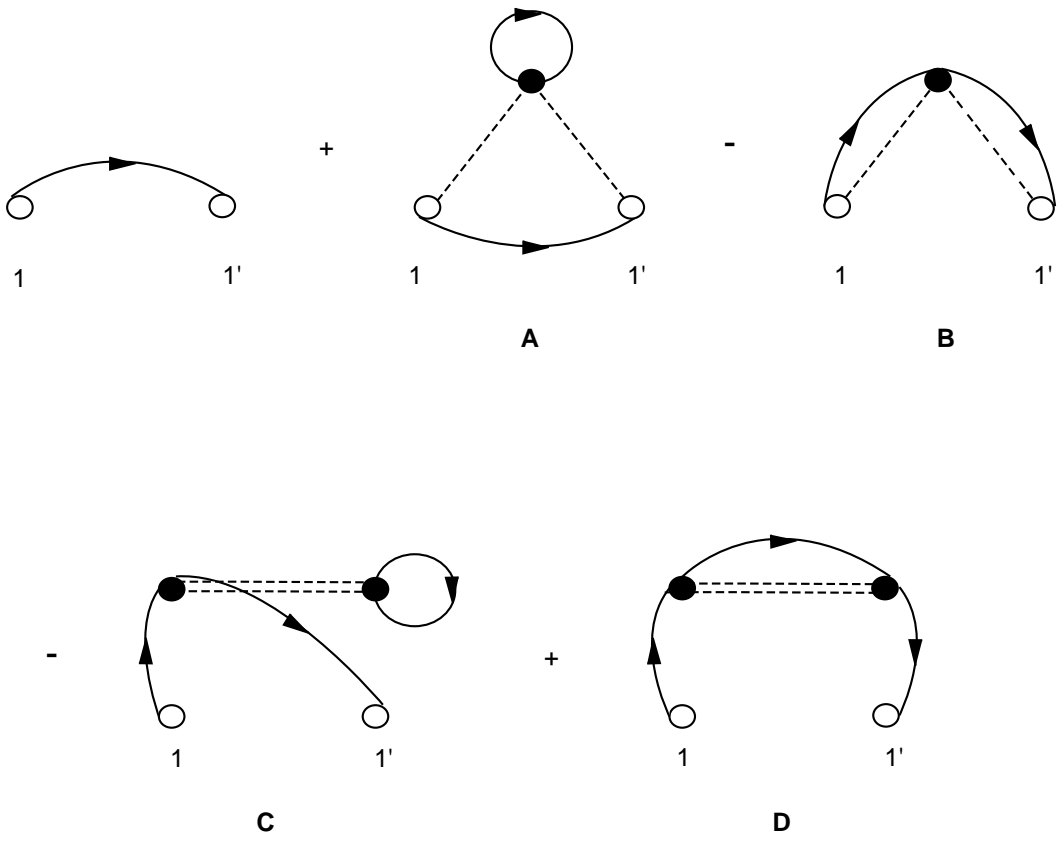
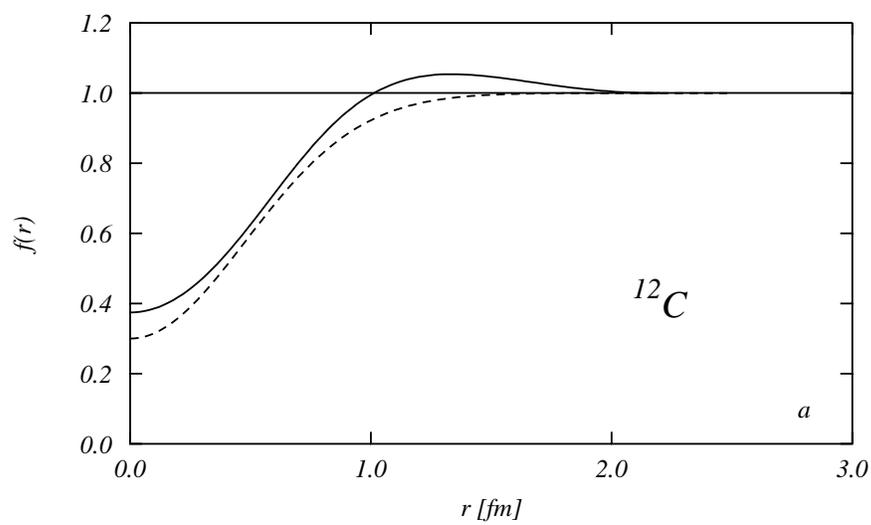
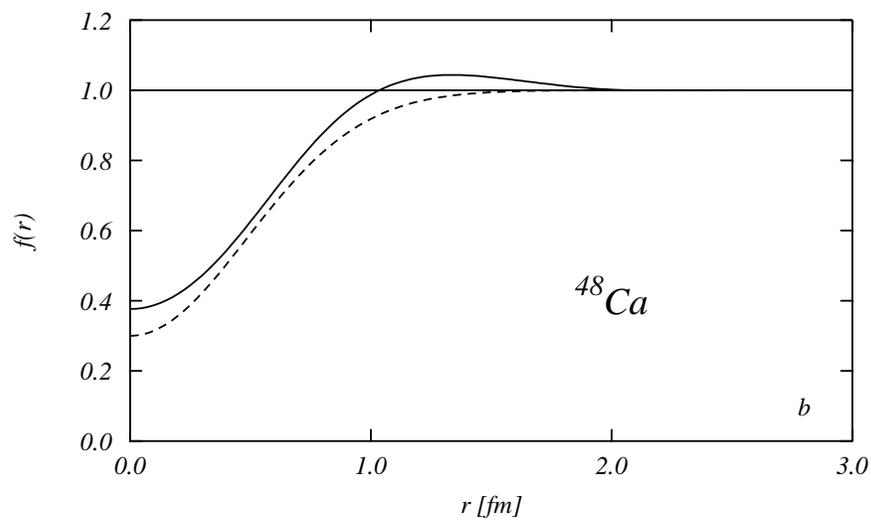
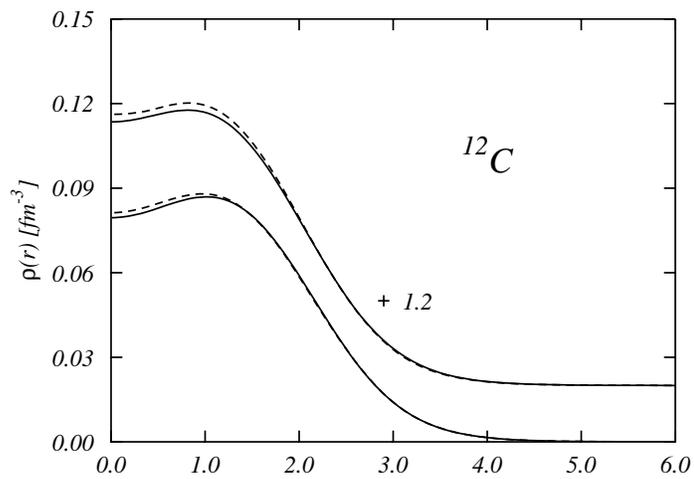
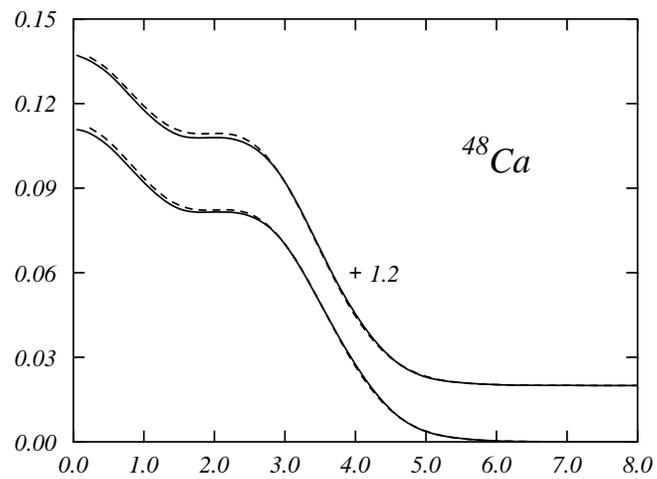
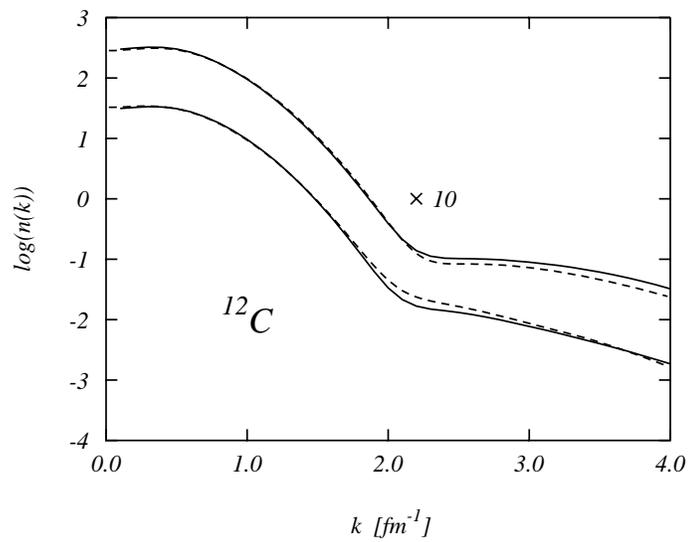
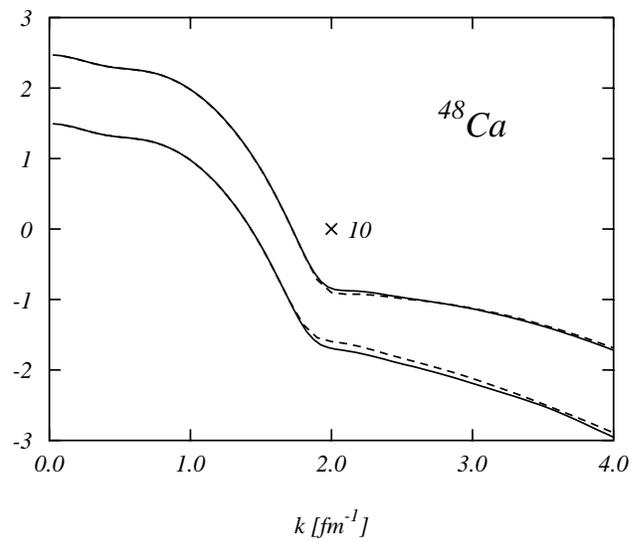
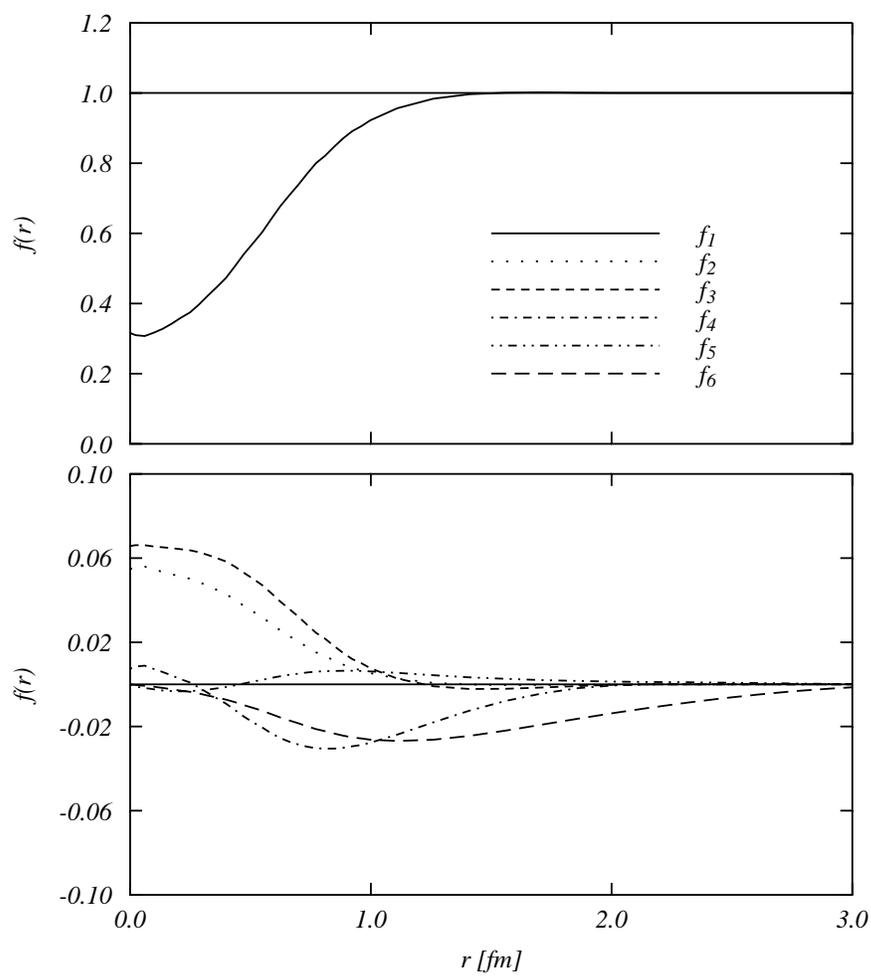
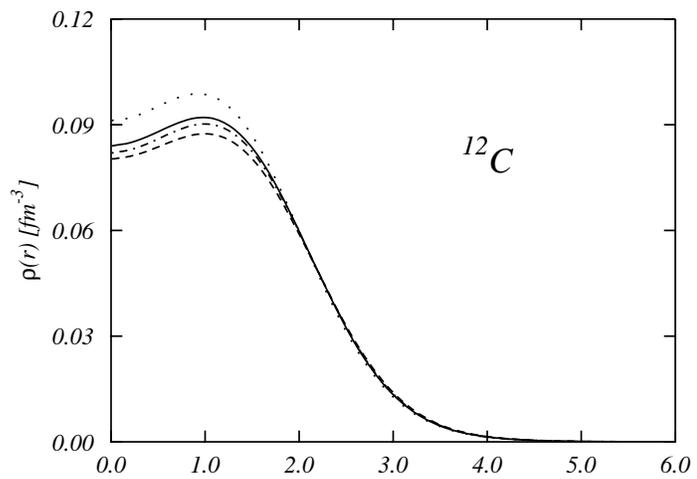
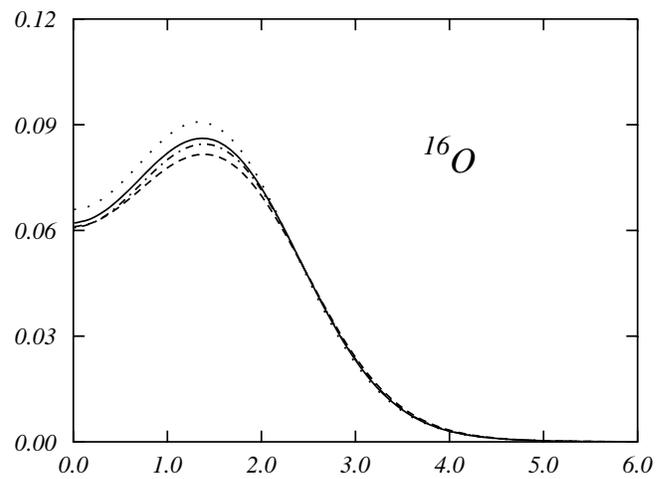
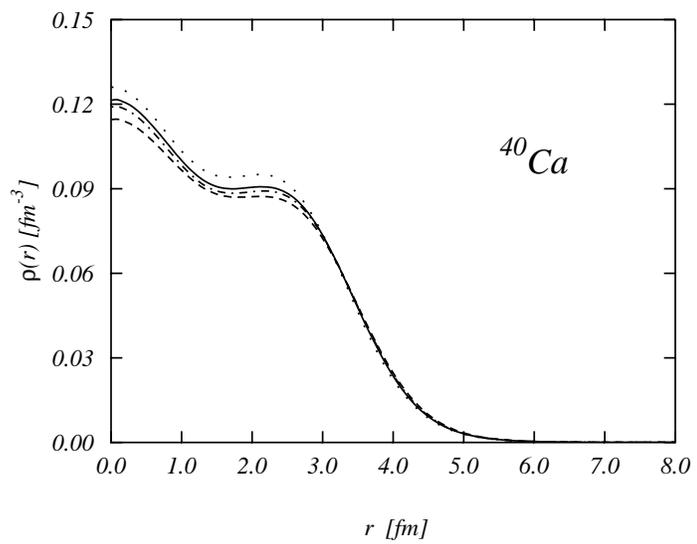
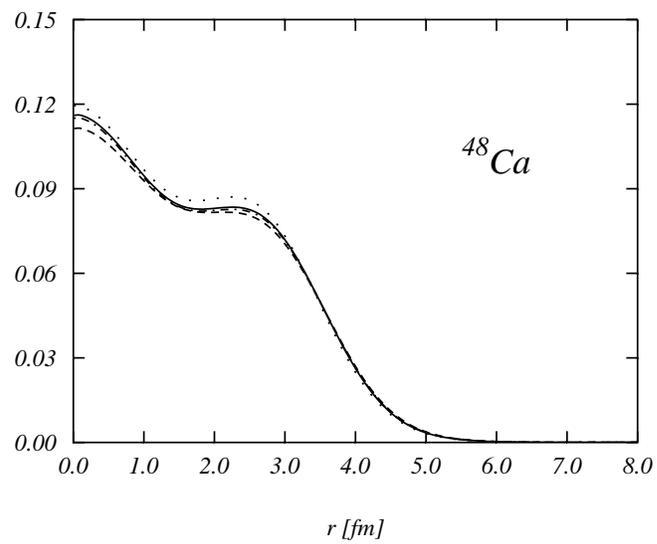
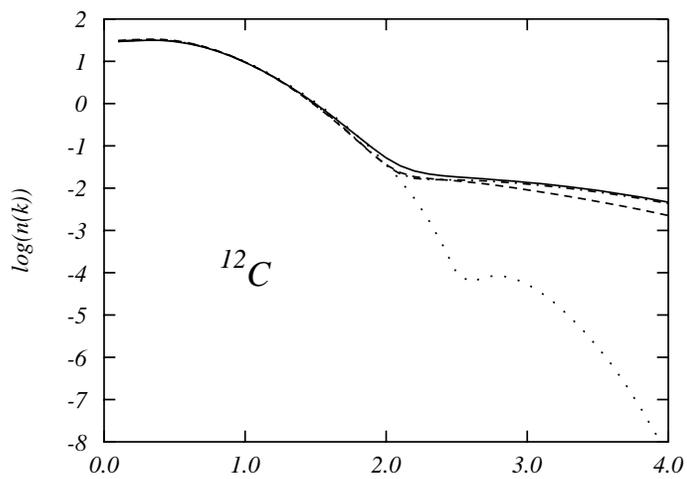
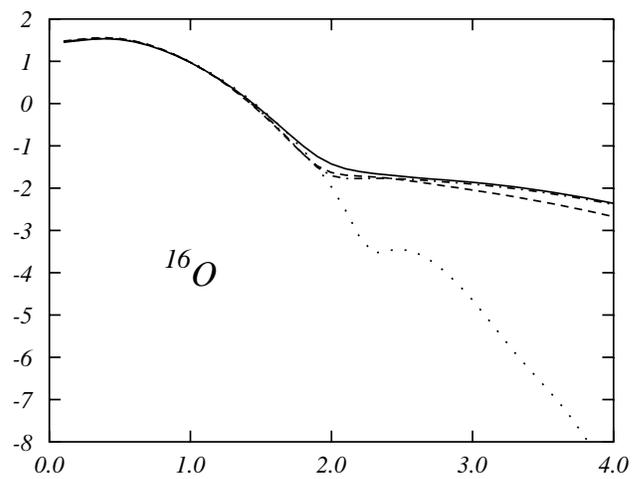
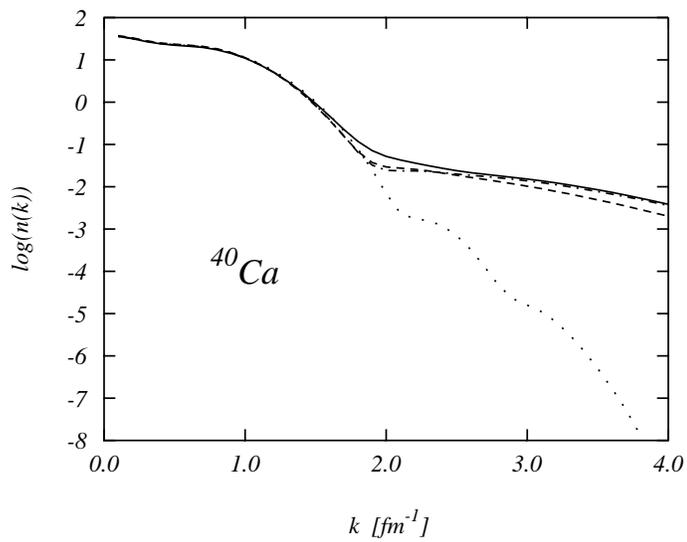
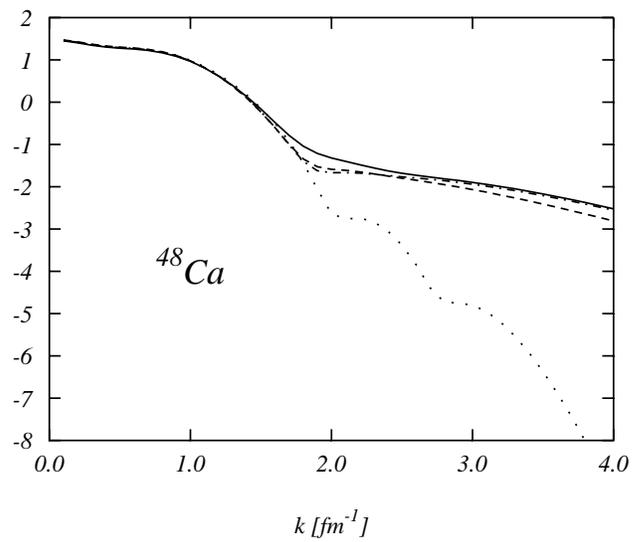
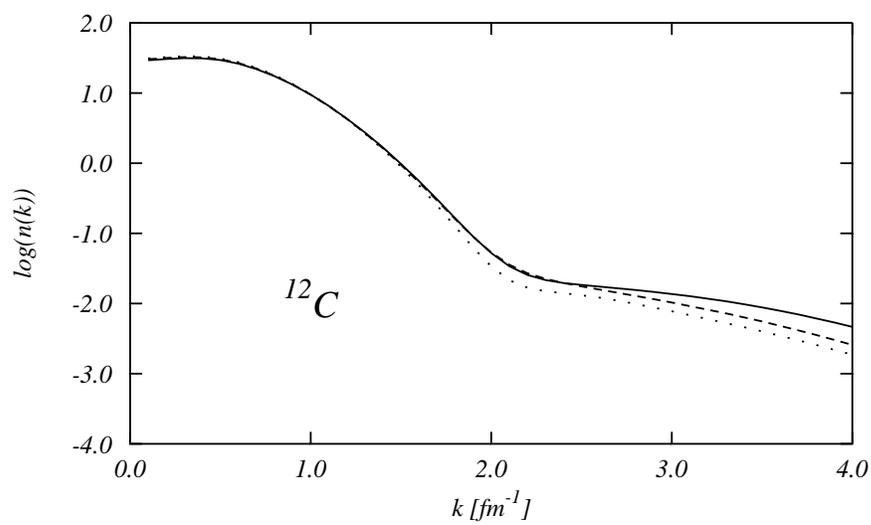
\end{document}